\documentclass{aastex}          
\usepackage{spr-astr-addons}    
\usepackage{url}

\usepackage{hyperref}

\newcommand{\kms}{\,km\,s$^{-1}$} 
\usepackage{amsmath, amsfonts}

\begin{document}
%
\title{Synthetic observations using POLARIS: an application to simulations of massive prestellar cores}

\shorttitle{Synthetic observations with POLARIS}
\shortauthors{Zamponi et al.}

\author{Joaquin Zamponi\altaffilmark{1,2}} 
\authoremail{jzamponi@udec.cl} 
\and 
\author{Andrea Giannetti\altaffilmark{3}}
\and 
\author{Stefano Bovino\altaffilmark{1}}
\and 
\author{Giovanni Sabatini\altaffilmark{1,3,4}}
\and 
\author{Dominik R. G. Schleicher\altaffilmark{1}}
\and 
\author{Bastian K\"ortgen\altaffilmark{5}}
\and 
\author{Stefan Reissl\altaffilmark{6}}
\and 
\author{Sebastian Wolf\altaffilmark{7}}

\altaffiltext{1}{Departamento de Astronom\'ia, Universidad de Concepci\'on, Esteban Iturra s/n Barrio universitario, Casilla 160, Concepci\'on, Chile. \\Email: jzamponi@udec.cl}
\altaffiltext{2}{Max-Planck-Institut f\"ur Extraterrestrische Physik (MPE), D85748 Garching, Germany}
\altaffiltext{3}{INAF - Istituto di Radioastronomia - Italian node of the ALMA Regional Centre (It-ARC), Via Gobetti 101, I-40129 Bologna, Italy}
\altaffiltext{4}{Dipartimento di Fisica e Astronomia, Universitá degli Studi di Bologna, Via Gobetti 93/2, I-40129 Bologna, Italy}
\altaffiltext{5}{Hamburger Sternwarte, Universität Hamburg, Gojenbergsweg 112, D-21029 Hamburg, Germany}
\altaffiltext{6}{Heidelberg University, Center for Astronomy, Institute of Theoretical Astrophysics, Albert-Ueberle-Str. 2, 69120 Heidelberg, Germany}
\altaffiltext{7}{Institut für Theoretische Physik und Astrophysik, Christian-Albrechts-Universität zu Kiel, Leibnizstr. 15, 24118 Kiel, Germany}

\begin{abstract}
   Young massive stars are usually found embedded in dense and massive molecular clumps which are known for being highly obscured and distant. 
   During their formation process, the degree of deuteration can be used as a potential indicator of the very early formation stages. This is particularly effective when employing the abundance of H$_2$D$^+$. However, its low abundances and large distances make detections in massive sources hard to achieve. 
   We present an application of the radiative transfer code \textsc{POLARIS}, with the goal to test the observability of the ortho-H$_2$D$^+$ transition $1_{10}$-$1_{11}$ ($\sim$372.42\,GHz) using simulations of high-mass collapsing cores that include deuteration chemistry. 
   We analyzed an early and a late stage of the collapse of a 60 M$_{\odot}$ core, testing different source distances. For all cases, we generated synthetic single-dish and interferometric observations and studied the differences in both techniques.
   The column densities we derive are comparable to values reported for similar sources. These estimates depend on the extent over which they are averaged, and sources with compact emission they can be highly affected by beam dilution. Combined ALMA-ACA observations improve in signal-to-noise ratio and lead to better column density estimates as compared to ALMA alone. We confirm the feasibility to study ortho-H$_2$D$^+$ emission up to distances of $\sim$7\,kpc.
   We provide a proof-of-concept of our framework for synthetic observations and highlight its importance when comparing numerical simulations with real observations. This work also proves how relevant it is to combine single-dish and interferometric measurements to derive appropriate source column densities. 
\end{abstract}


%

\section{Introduction}
\label{sec:Introduction}
Massive stars, with masses above $\sim$8-10~M$_{\odot}$, significantly impact the environment in which they are born. They affect the thermal properties and chemical composition of the parent cloud via photoionization and dust heating onto the circumstellar discs of their neighbouring low-mass stars. They end their lives as supernovae, impacting the surroundings and the subsequent star formation by depositing high amounts of heavy elements and increasing the level of turbulence within the cloud. 
Different scenarios have been proposed to explain the formation of massive stars, however, no global consensus has yet been found on how this occurs (see \citealt{McKeeAndOstriker,ZinneckerAndYorke07} or \citealt{MotteBontemsAndLouvet} for a review). 
\citet{Bonnell97,Bonnell01} proposed the competitive core accretion model, where all bound objects accrete gas from their surroundings; objects placed in the center of the cloud become more massive than in the outskirts, thanks to the favourable conditions for accretion at the bottom of the potential well. 
\citet{McKeeAndTan02,McKeeAndTan03} proposed a scaled-up version of the low-mass star paradigm \citep{Shu87}, termed turbulent core accretion model, where massive prestellar cores are supposed to have high gas pressures, supersonic turbulence and significant magnetic support, leading to a rather slow, almost monolithic and unfragmented collapse.
One of the largest differences between these two models is the predicted collapse timescale. 
Therefore, a proper analysis of the collapse evolution is needed to distinguish between different formation scenarios.
A useful observational tool to measure such timescales are chemical clocks, i.e., molecular tracers that show drastic abundance changes with density and temperature variations as a function of time (e.g., \citealt{Beuthe09, Fontani11_proceedings, Molinari16, Urquhart19, Coletta20, Sabatini21, Mininni21}).
To exploit this technique, it is possible to combine the information from different tracers. 
However, this is highly dependent on the observability of each molecule transition. 

Infrared Dark Clouds (IRDC), massive quiescent clouds that represent the most likely birthplaces for the next generation of high-mass stars, are characterized by low gas temperatures ($T_{\rm gas}<$20K), high gas column densities ($N_{\rm gas}\sim10^{23-25}$ cm$^{-2}$) and a large degree of CO-depletion~(e.g., \citealt{Fontani11,CaselliAndCeccarelli12,Caselli13,Giannetti19,Jorgensen20,Sabatini20}; see also \citealt{BerginAndTafalla07} for a review). 
The absence of C-bearing molecules in the gas phase enables further chemical reactions to take place, such as the enhancement of the abundances of deuterated molecules. 
Deuteration reactions start from the proton-deuteron exchange between HD and H$_3^+$, 
\begin{eqnarray}
{\rm H}_3^+ +{\rm HD}&\rightleftharpoons&{\rm H}_2{\rm D}^+ +{\rm H}_2 +\Delta E_1,\\
{\rm H}_2{\rm D}^+ +{\rm HD}&\rightleftharpoons&{\rm D}_2{\rm H}^+ +{\rm H}_2 +\Delta E_2,\\
{\rm D}_2{\rm H}^+ +{\rm HD}&\rightleftharpoons&{\rm D}_3^+ +{\rm H}_2 +\Delta E_3,
\end{eqnarray}
where $\Delta E_1$, $\Delta E_2$ and $\Delta E_3$ depend on the isomers involved in each reaction~\citep{Hugo07}. 
This set of reactions increases the abundances of deuterated species such as H$_2$D$^+$ and D$_2$H$^+$ over time. 
However, they can also be efficiently destroyed by the presence of CO in the gas phase, via
\begin{eqnarray}
{\rm H}_3^+ +{\rm CO}&\rightleftharpoons&\rm{HCO}^+ +{\rm H}_2,\\
{\rm H}_2{\rm D}^+ +{\rm CO}&\rightleftharpoons&{\rm DCO}^+ +{\rm H}_2. \label{eq:h2d+_to_dco+}
\end{eqnarray}
In the cold and dense inner regions of IRDCs, the absence of CO boosts the formation of deuterated species and provides an insight into the onset of collapse in prestellar cores~\citep{Dalgarno&Lepp84}. 
Models of the evolution of such deuteration reactions \citep{Walmsley04,van_der_Tak05,Flower06,Sipila13,Bovino19} show a maximum H$_2$D$^+$ abundance right before the formation of a protostellar object, which would then heat up the gas and evaporate the CO from the dust grain surfaces, converting H$_2$D$^+$ back into DCO$^+$ (following  reaction~\ref{eq:h2d+_to_dco+}), and decreasing the abundance of H$_2$D$^+$.

All the models that attempt to estimate the evolution of deuteration reactions are subject to the initial value of the ratio between the different isomers of H$_2$ (ortho-to-para ratio). This ratio can be observationally traced using the  H$_2$D$^+$ and D$_2$H$^+$ molecules as proxies (e.g., \citealt{Flower06,Hugo09,Brunken14,Bovino21}). 
Unfortunately, observations of low energy transitions from para-H$_2$D$^+$ suffer from large atmospheric attenuation at terahertz frequencies, hence, most of the efforts have been directed to observe the sub-millimeter transitions of ortho-H$_2$D$^+$ $J_{\rm {K_a, K_c}} = 1_{10}$-$1_{11}$ at $\sim$372.42~GHz \citep{AmanoAndHirao05}. 
Several detections have been reported toward low-mass cores \citep{Caselli03,Vastel06,Caselli08,Parise11,Friesen14,Brunken14,Miettinen20}, since they are relatively close and are therefore easier to detect. 
In the context of high-mass sources, single-dish observations have also been carried out  \citep{Harju06,Swift09,Pillai12,Giannetti19,Sabatini20}. 
However, massive sources are rather distant ($\gtrsim$1\,kpc; \citealt{ZinneckerAndYorke07, Giannetti14} and \citealt{Konig17}), and the angular resolutions available are not sufficient to resolve the inner regions of the cores. 
\citet{Sabatini20} detected o-H$_2$D$^+$ toward 16 high-mass star-forming regions in different evolutionary stages, observed with the \textit{Atacama Pathfinder EXperiment} 12-m telescope (APEX; \citealt{Gusten06}), and found a correlation between the evolutionary state of the clumps and the abundance of o-H$_2$D$^+$, which is higher in younger sources (see also \citealt{Giannetti19}).  
More recently, \citet{Redaelli21} presented the first interferometric detections of o-H$_2$D$^+$ in high-mass star forming clumps, observed with the \textit{Atacama Large Millimeter/submillimeter Array} (ALMA; \citealt{Wootten09}). 
Their observed line emission of o-H$_2$D$^+$ is rather narrow and subsonic, suggesting that the gas in their sources is cold ($>1$0\,K) and dense ($>$ 10$^6$\,cm$^{-3}$), indicating the lack of protostellar heating and more representative of a prestellar and sub-virial phase. 

In this work, we developed a framework for synthetic observations and we applied it to a set of numerical simulation of massive star formation, using the observability of the o-H$_2$D$^+$ transition $J_{\rm {K_a, K_c}} = 1_{10}$-$1_{11}$ as our case of study. 
Our aim is to show the difficulties involved in the observations of such distant and obscure sources.  
For this, we performed a set of radiative transfer (RT) simulations using these simulated cores as synthetic sources (see~\ref{subsec:Synthetic_sources}) and then post-processed them by adding instrument-related effects. 
In this last step, we distinguished between single-dish and interferometric observations, looking for an understanding of the key differences that may arise when observing the same source with both techniques. 
Finally, we derived column densities of o-H$_2$D$^+$ from the resulting intensity distributions and compared our results both to values reported in the literature and also to the physical column densities measured from the model. 

In section~\ref{sec:Methods} we describe the steps followed for each synthetic observation.
In section~\ref{sec:Results} we show the radial distributions of the synthetic maps and the column densities derived and then we present a comparison of the column densities derived from a single-dish and a interferometer.  
In section~\ref{sec:Discussion} we discuss the limiting cases of our results in terms of source distance and observing time. 
Finally, we provide a summary and conclusions in section~\ref{sec:Summary_and_conclusion}.


\begin{figure}
    \includegraphics[width=\columnwidth,trim=7cm 2.3cm 7cm 2.3cm, clip]{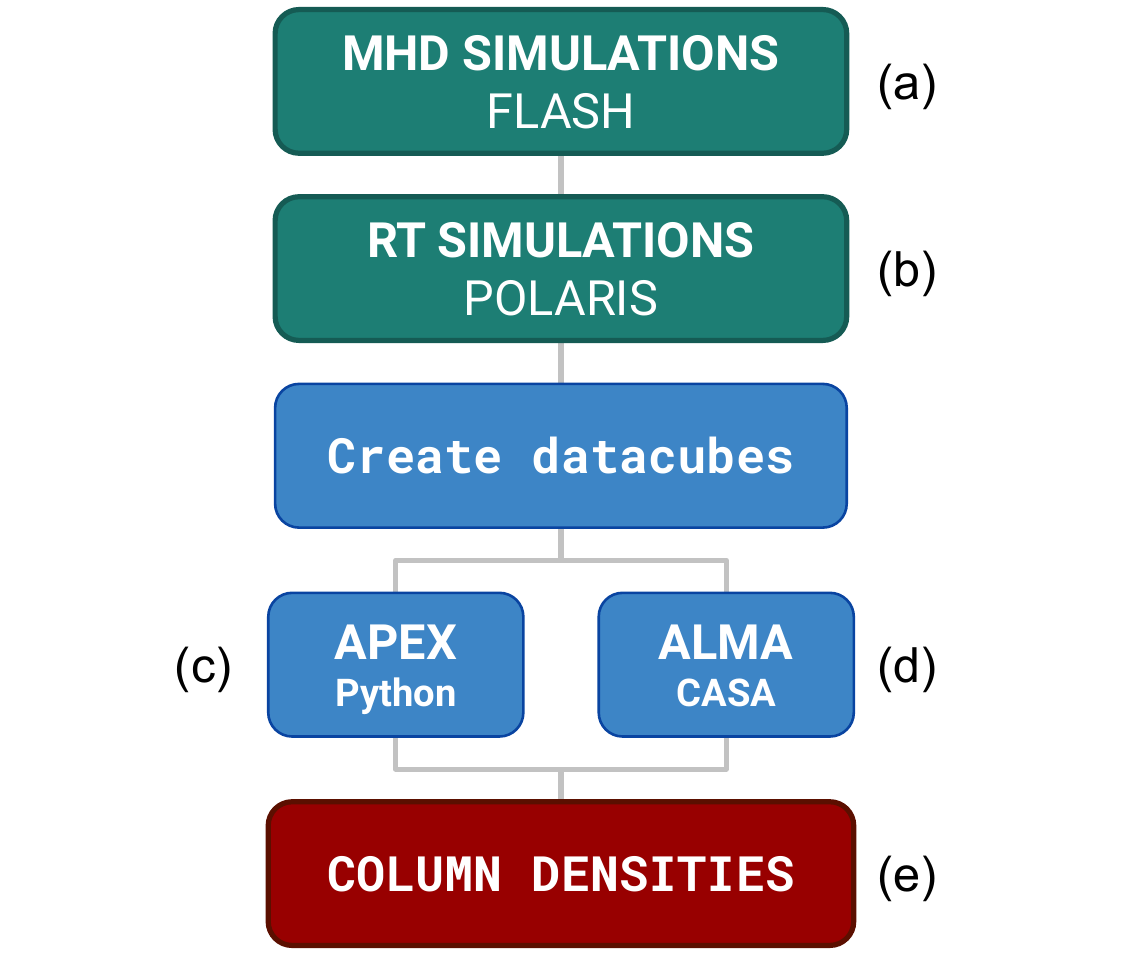}
    \caption{Workflow of each synthetic observation. The input MHD simulations are ray-traced using the POLARIS RT code and then post-processed to distinguish between interferometric and single-dish simulations, using the CASA software and a Python module written for this project, respectively. Each resulting intensity distribution is then converted into column densities for further comparison with real data.}
    \label{fig:workflow}
\end{figure}

\begin{table*}
    \caption[Initial parameters of the cores selected]{Initial parameters of the core selected from~\citet{Kortgen17}, labeled Lmu10M2. The name refers to a source within their sample that has a low surface density, a mass-to-flux ratio of 10 and a Mach number of 2.}
    \label{tab:initial_conditions}
    \centering
    \setlength\tabcolsep{6pt} 
    \begin{tabular}{l c c c c c c c c }
    \hline
     & Surface & Core & Core & Av. field & Mass-to- & Mach & Virial & Free-fall \\
    Run & density  & radius & mass & strength & flux ratio & number & parameter & time \\
      & (g cm$^{-2}$) & (pc) & (M$_{\odot}$) & ($\mu$G) & $\mu/\mu_{\rm crit}$ & $\mathcal{M}_{\rm turb}$ & $\alpha_{\rm vir}$ & (kyr)\\ 
    \hline
    Lmu10M2 & 0.14 & 0.17 & 60 & 27 & 10 & 2 & 0.64 & 149 \\
    \hline
    \end{tabular}
\end{table*}

\begin{figure*}
    \centering
    \includegraphics[ width=17cm]{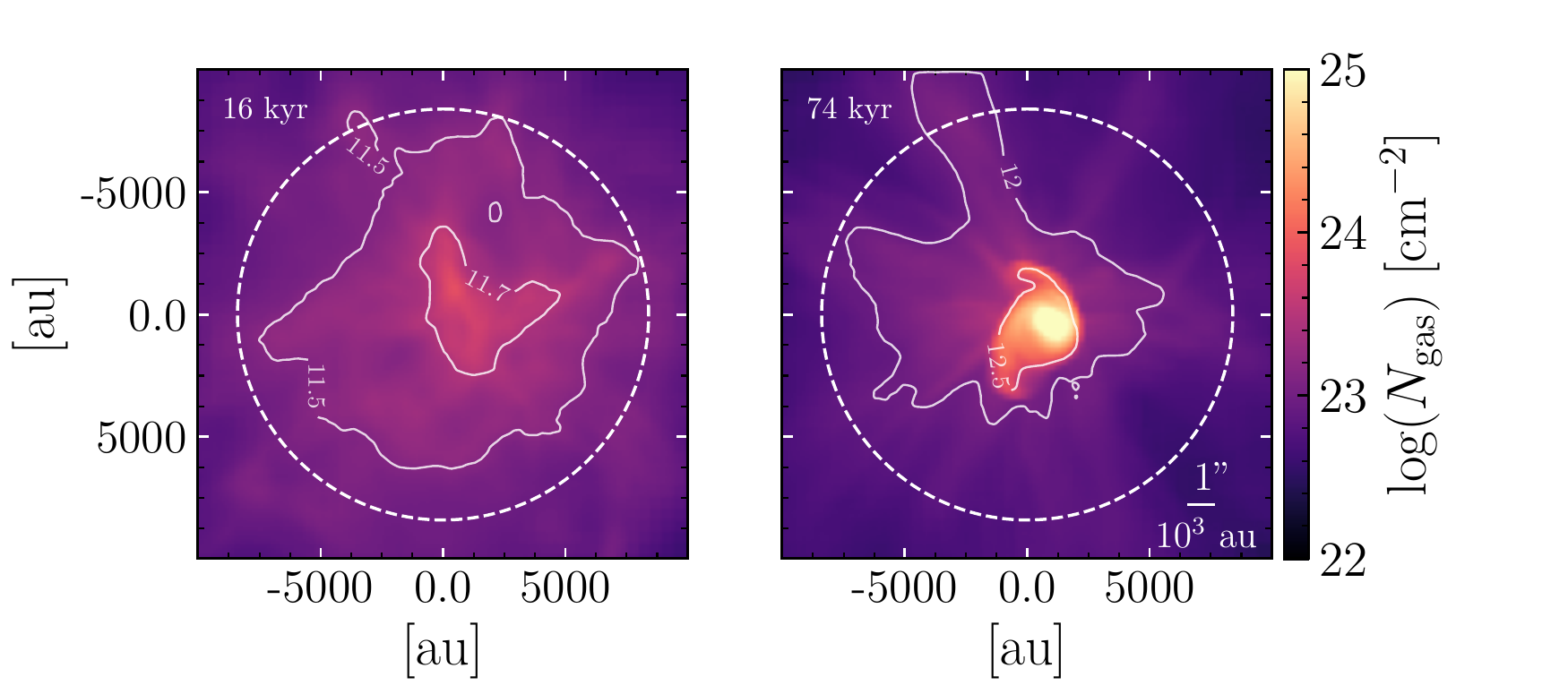}
    \caption{Time evolution of the gas column density in the simulated collapsing core, shown at the two timesteps analyzed in this work. White contours represent the column density of o-H$_2$D$^+$ and the dashed circle represents the field of view of the APEX and ALMA telescopes (16.8" at 372.42\,GHz) at a distance of 1\,kpc. The scalebar of 1" illustrates the synthesized beam of the ALMA synthetic observations of the compact source.}
    \label{fig:mhd_snapshots}
\end{figure*}

\begin{figure}
    \centering
    \includegraphics[width=\columnwidth]{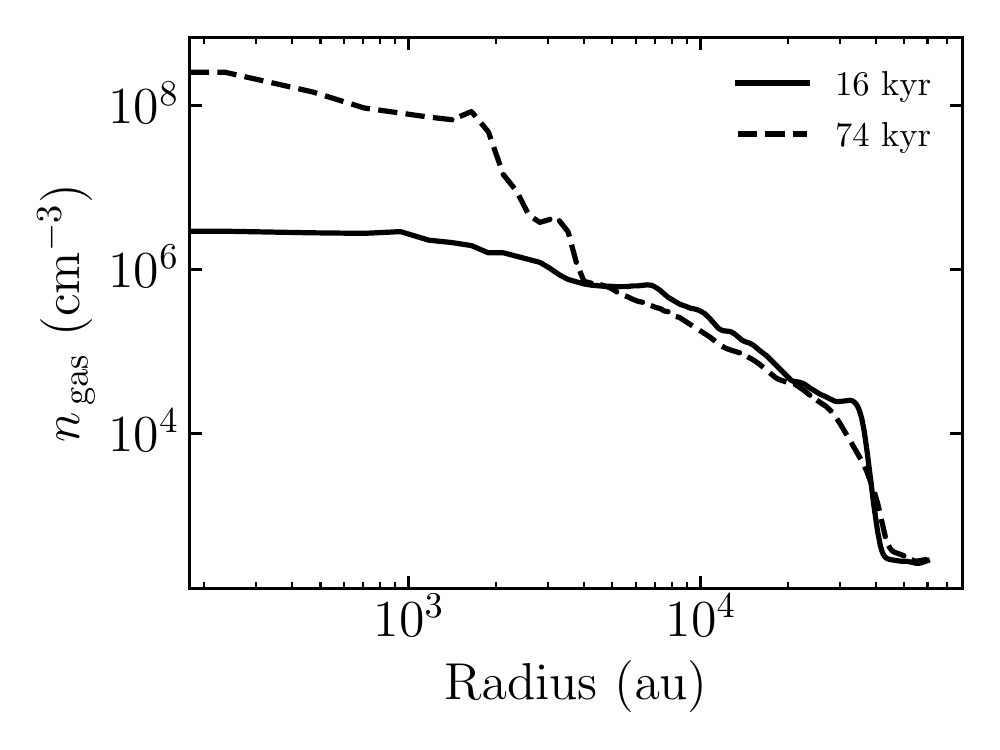}
    \caption{Radially averaged midplane density profiles for the two time snapshots shown in Fig.~\ref{fig:mhd_snapshots}. }
    \label{fig:radial_density_profiles}
\end{figure}

\section{Methods}
\label{sec:Methods}
\subsection{Workflow}
\label{subsec:Workflow}

To perform a successful synthetic observation, three main ingredients are needed: a synthetic source, a ray-tracer and a synthetic detector.
The synthetic source, or model, can be either a simple density and temperature distribution or a full three dimensional magneto-hydrodynamic (MHD) simulation~\citep{Haworth18} (Fig.~\ref{fig:workflow}a). 
The ray-tracing is done by a RT code, which calculates the propagation of light within the source, accounting for emission, absorption and scattering, and creates a resulting flux distribution attenuated at a given distance (Fig.~\ref{fig:workflow}b).
The synthetic detector is the numerical array in which this ideal intensity distribution is stored.
To these intensity maps, instrument-related effects must be applied, such as convolution with a telescope beam (in the case of single-dish telescopes) or a reconstructed beam (in the case of inteferometers), the addition of noise and image reconstruction from a complex visibility, in the case of interferometers (Fig.~\ref{fig:workflow}c,d). 
Thus, in order to make synthetic observations as realistic as possible, the intensity maps from the RT simulations must be post-processed based on the specific properties of the telescope of interest.
The fluxes resulting from each synthetic observation are finally converted into column densities, to then be compared against the values from the numerical  simulations  (Fig.\ref{fig:workflow}e).
All the functions and routines used in steps (c), (d) and (e) from Fig.\ref{fig:workflow} are provided in an online repository\footnote{\url{https://github.com/jzamponi/synthetic_module}}.

\subsection{Synthetic source}
\label{subsec:Synthetic_sources}
The synthetic source used in this work is an isolated magnetized massive prestellar core taken from the set of 3D ideal-MHD simulations performed by~\citet{Kortgen17}.
They employed the \textsc{Flash} code (v4.2.2)~\citep{FLASH}, coupled with the \textsc{Krome} package~\citep{Grassi14} to follow the deuteration chemistry of light hydrides.

\subsubsection{Initial conditions}
\label{subsubsec:Initial_conditions}
The core is initialized as an isolated Bonnor-Ebert (BE) sphere~(\citealt{Bonnor56} and \citealt{Ebert95}), supersonically turbulent and assumed to collapse isothermally at $T_{\rm gas}=15$~K. 
The initial conditions of the core are listed in Table~\ref{tab:initial_conditions}. 
The mass of the core is 60\,M$_{\odot}$. 
The central gas number densities evolve from $n_{\rm gas}\sim 3\times10^6$~cm$^{-3}$ ($N_{\rm gas}=10^{23.5}$~cm$^{-2}$; at 0.1~free-fall times) to $\sim 1\times10^8$~cm$^{-3}$ ($N_{\rm gas}=10^{25}$~cm$^{-2}$; at 0.5~free-fall times) as seen in Fig. \ref{fig:mhd_snapshots}. 
Radially averaged density profiles along the midplane of the core are also shown in Fig.~\ref{fig:radial_density_profiles}, which show that the core is indeed a compact source with most of the dense material lying within the inner $\sim$6000\,au.  

\subsubsection{Deuteration chemistry}
\label{subsubsec:Deuteration_chemistry}
The simulations from~\citet{Kortgen17} included a detailed non-equilibrium chemical network, with 21 chemical species in the gas phase and also dust grains, along with their ionized states. 

The network solved the deuteration reactions based on~\citet{Walmsley04}, who assumed full depletion of elements heavier than He, which accounts for the CO freeze-out onto dust icy mantles that has been observed to be effective at the densities studied here ($\gtrsim10^4$~cm$^{-3}$; \citealt{Caselli99, Tafalla02, Giannetti14}). This assumption allows to significantly reduce the network and the computational time of the simulation, while still providing realistic results~(e.g. \citealt{Sabatini19} and \citealt{Bovino19}). 
Electron attachment, recombination of positive ions and grain surface reactions were also included, such as the formation of H$_2$ and HD, with the exception of D$_2$, which is mainly formed in the gas-phase.
The full chemical network was solved and evolved using the \textsc{Krome} package \citep{Grassi14}.

\subsection{Radiative transfer calculations}
\label{subsec:Radiative_transfer_calculations}
We performed ray-tracing RT simulations with the code \textsc{Polaris}\footnote{\url{http://www1.astrophysik.uni-kiel.de/~polaris}} (v4.06)~(\citealt{Reissl16} and \citealt{Brauer17}) based on the temperature and density distribution taken from two timesteps of the collapsing core simulations, starting from slightly after the initial state (0.1~$t_{\rm ff}$; with $t_{\rm ff}$ the free-fall time, 16\,kyr) until 0.5~$t_{\rm ff}$ (74\,kyr, see Fig.~\ref{fig:mhd_snapshots}).
We simulated the emission of the o-H$_2$D$^+$ transition $J_{\rm {K_a, K_c}} = 1_{10}$-$1_{11}$ at $\sim$372.42~GHz.  
For simplicity, dust was not included in our setup, to reduce the uncertainty in the continuum subtraction from the spectra and assuming optically thin emission for the o-H$_2$D$^+$. 
This assumption was assessed by computing the line optical depth from the synthetic cubes, using equation \ref{eq:optical_depth}. 
We obtained values no larger than 0.1 for even the most dense stages of the core, as for instance that shown in Fig. \ref{fig:mom0_oh2dp}. 
However, the exclusion of a dust component in the setup can only be done safely when the opacity from the dust is known to be low enough so that line emission is not significantly extincted. 
In order to quantify the effect, we assume a dust composition of silicates and graphites, previously used in the modelling of dust in pre- and proto-stellar cores \citep{Draine84, OssenkopfHenning1994}. 
This composition, at the frequency of 372.42~GHz, yields an opacity of $\kappa_{\rm abs}<4.56$~cm$^2$g$^{-1}$. This means that only in the peak  dust surface density of the evolved stage of the core ($\sim1$~g cm$^{-2}$; for a dust-to-gas mass ratio of 0.01), the optical depth would be $\tau_{\nu}\sim4.56$, and much lower around it. 
This upper limit in the opacity is based on the dust coagulation model of  \citet{OssenkopfHenning1994} after 100\,kyrs of evolution, which is longer than the latest time in our collapsing core simulations. 
The opacity they provide for the initial state of the coagulation model yields $\tau\sim0.9$, which means that the emission from both of the collapse stages we considered should be marginally optically thick in most of the core scales, with the only exception of the peak surface density at 74\,kyr. Based on this, we neglect the effects of dust opacity in this work but acknowledge that it should be important to take into account in more detailed investigations.

We studied distances from 1 up to 10 kpc. 
For each time and distance we generated spectral cubes, covering \mbox{10 km s$^{-1}$}  centered on the rest frequency of our line of interest, i.e. $\sim$372.42~GHz. 
We fixed the spectral resolution of each data\-cube to 0.03~km\,s$^{-1}$ ($\sim$38~kHz) for all simulations, by splitting the spectral range into 333 channel maps, meant to represent the highest resolution offered by the FLASH$^+$ dual-frequency MPIfR principal investigator (PI) receiver~\citep{FLASH+} at a frequency of $\sim$372.42~GHz, mounted at the APEX telescope.   
Based on the gas column density distribution of the core (see Fig.~\ref{fig:mhd_snapshots}), the lowest number densities are around 1.62$\times$10$^5$~cm$^{-3}$, for a core radius of 0.2~pc.
Such number densities are above the critical density of the o-H$_2$D$^+$ line emission at 372.42~GHz  (1.3$\times$10$^5$~cm$^{-3}$; \citealt{Hugo09}) and therefore Local Thermodynamic Equilibrium (LTE) can be assumed when computing the level populations.  
The assumption of LTE means that the excitation temperature $T_{\rm ex}$ and the kinetic gas temperature $T_{\rm gas}$ are  equal (15~K), and then 
\begin{equation}
    T_{\rm gas}=T_{\rm ex}=\frac{h\nu_{ij}}{k_{\rm B}}\left[\ln{\left(\frac{g_i\,n_j}{g_j\,n_i}\right)}\right]^{-1},
\end{equation}
with $g_i$ and $g_j$ the statistical weights, and  $n_i$ and $n_j$ the level populations following the Boltzmann distribution.
The molecular data of o-H$_2$D$^+$ used in our simulations was obtained from the Cologne Database for Molecular Spectroscopy\footnote[2]{\url{https://cdms.astro.uni-koeln.de/cdms/portal}} (CDMS;~\citealt{Muller2005,Endres2016}).

\subsection{Single-dish observations}
\label{subsec:APEX_simulations}
The single-dish synthetic observations were performed as a post-processing of the ideal intensity maps from the radiative transfer calculation. This was done by adding instrument related effects to the datacubes, such as the convolution of the cubes with a telescope beam and the addition of thermal noise (Fig.~\ref{fig:workflow}c).

\subsubsection{PSF convolution}
We convolved the datacubes with a 2D Gaussian kernel, resembling the Point-Spread-Function (PSF) of a parabolic single-dish telescope.
We used a Full-Width-at-Half-Maximum (FWHM) beam size of 16.8 arcseconds, that is the effective resolution achieved by the APEX 12-m dish at a frequency of $\sim$372.42~GHz. 
After the convolution, we converted the maps from Jy\,pixel$^{-1}$ into Jy\,beam$^{-1}$ by rescaling the flux with the ratio of the area of a gaussian beam over the area of a square pixel, as
\begin{equation}
  \frac{F}{{\rm Jy\,beam}^{-1}} = \frac{\pi}{4\ln 2}\frac{\theta_{\rm maj}\,\theta_{\rm min}}{{\rm arcsec}^2}\left(\frac{\rm pix\,size}{\rm arcsec}\right)^{-2} \frac{F}{{\rm Jy\,pixel}^{-1}},
\end{equation}
where $\theta_{\rm maj}$ and $\theta_{\rm min}$ are the FWHM of the major and minor axis of the beam in arcseconds, equal for a circular beam.
At a distance of 1~kpc, the APEX-beam corresponds to 16800~au.
This means that these observations were spatially unresolved almost during the entire evolution.

Convolution onto the images was performed using the \texttt{convolve\_fft}\footnote{\url{https://docs.astropy.org/en/stable/convolution}} function from the Astropy Python package~\citep{astropy2013,astropy2018}.
This function performs a Fourier-space convolution of 2D data-matrix with a Gaussian kernel of standard deviation $\sigma={\rm FWHM}\,/\,\sqrt{8\ln 2\,}$, where FWHM is the angular resolution in number of pixels.

\subsubsection{Addition of noise}
We converted the images from Jy beam$^{-1}$ to brightness temperature ($T_{\rm b}$) and added Gaussian noise to them, with a standard deviation ($T_{\rm rms}$) derived from the equation 
\begin{equation}
    \label{eq:T_rms_single-dish}
    T_{\rm rms} = \frac{T_{\rm sys}}{\sqrt{\Delta\nu\,t_{\rm int}}}, 
\end{equation}
where $\Delta\nu$ is the spectral resolution, $t_{\rm int}$ is the integration time and $T_{\rm sys}$ is the system temperature~\citep{Kraus_Radio_Astronomy}.
To obtain realistic values of $T_{\rm sys}$ for a given observing setup, we used the APEX Observing time calculator\footnote{\url{http://www.apex-telescope.org/heterodyne/calculator}}.
The noise level added to the simulations was $T_{\rm rms}\sim8.6$~mK, corresponding to $T_{\rm sys}\sim543$~K for an integration time on source of 6~hrs, a source elevation of 45~deg, a precipitable-water-vapor (pwv) level of 0.5~mm (typical for APEX observations at 372.42~GHz, e.g. \citealt{Miettinen20,Sabatini20}) and a spectral resolution of $\Delta v=0.3$~km\,s$^{-1}$, achieved by binning the spectra with a factor of 10. 
We emphasize that the integration times presented here and in section \ref{subsec:min_int_time} correspond to on-source times only and do not consider calibration observations.  

\subsection{Interferometric observations}
\label{subsec:ALMA_simulations}
We performed interferometric synthetic observations for ALMA, to ease the comparison with single-dish observations since it represents an array of APEX-like telescopes.  
We simulated the interferometric response of our intensity maps by using the 
\textit{Common Astronomy Software Applications package} (CASA\footnote{\url{https://casa.nrao.edu/}}) tasks \texttt{simobserve}\footnote{\url{https://casa.nrao.edu/docs/taskref/simobserve-task.html}} and \texttt{tclean}\footnote{\url{https://casa.nrao.edu/docs/taskref/tclean-task.html}}.
\texttt{simobserve} is used to create a visibility measurement set from an image model and \texttt{tclean} to reconstruct an image out of the visibility table using the \textsc{clean} algorithm.
An example script for the ALMA simulations is provided in the online repository mentioned in section~\ref{fig:workflow}.
The input image models were the output from the radiative transfer calculations, in units of Jy\,pixel$^{-1}$. 

We used the most compact array configuration C43-1 in band 7 (Cycle 7), obtaining a synthesized beam of about $1".01\times0".92$ at 372.42~GHz, in order to achieve the highest sensitivity among all configurations and to recover most of the extended emission. 

Part of the analysis included simulations using the \textit{Atacama Compact Array} (ACA;~\citealt{Iguchi09}). 
ACA is used to patch the shorter baselines in the \textit{uv}-plane, not covered by the ALMA main array, retrieving information from large scale emission. 

This configuration also aids in reducing the spatial filtering of the extended emission.
Based on the same argument, we cleaned with natural weighting for all images, aiming for improved sensitivity over angular resolution, since the source is always resolved in our interferometric observations. 
The number of pixels and image size were adjusted accordingly for every source distance that we studied, in order to ensure a Nyquist sampling of the image (i.e., 5 pixels across the beam minor axis).   
We cleaned interactively, to define a proper mask for the source at each timestep and distance, checking the residual map at each major cycle. 
The cleaning threshold was also adjusted interactively and a multiscale deconvolver was used, for scales of 0 (point-source), 1, 5 and 10 times the beam size.   
The gridder was set to \textit{standard}. 
The spectral channel width was 0.03~\kms, although we also binned the spectrum up to 0.3 and 1~\kms \, when analyzing the low-S/N observations at large source distances. 
Thermal noise was added by the \texttt{simobserve} task, based on a sky temperature of 114.3~K, atmospheric zenit opacity ($\tau_0$) of 0.579, precipitable-water-vapor of 0.658 and a ground temperature of 269~K.
This led to noise levels of $\sim$1.4~mJy beam$^{-1}$ (9~mK) for a 10~hrs integration time.

\subsection{Derivation of column densities}
\label{subsec:Derivation_of_column_densities}

\label{subsubsec:The_optical_depth}
We derived source column densities from the single-dish and inteferometric synthetic fluxes, following the procedure described in \citet{MagnumAndShirley}, and similarly used in \citet{Vastel06, Caselli03, Busquet10} and \citealt{Parise11} for this specific line, where the total column density ($N$) of a given species X is given by 
\begin{equation}
    \label{eq:column_density}
    \frac{N(\mathrm{H_2D^+})}{\rm cm^{-2}} = \frac{8\pi\nu^3}{c^3}\frac{Q(T_{\rm ex})}{g_{u}A_{ul}}\frac{e^{E_{u}/T_{\rm ex}}}{e^{h\nu/k T_{\rm ex}}-1}\int\tau dv,
\end{equation}
with $u$ and $l$ referring to the upper and lower levels of the transition, respectively,
$k$ and $h$ are the Boltzmann and Planck constants, respectively and $\nu=372.42$~GHz is the frequency of the transition $J_{\rm {K_a, K_c}} = 1_{10}$-$1_{11}$. The statistical weight of the upper level and the Einstein coefficient for the transition are $g_u=9$ and $A_{ul}=1.08\cdot10^{-4}$~s$^{-1}$, respectively. $E_u=17.87$~K is the energy of the upper level, $Q(T_{\rm ex})=11.70$ is the partition function of the molecule (for the ortho isomer only) at $T_{\rm ex}=15$~K and $\tau$ is the optical depth of the line, obtained from
\begin{equation}
    \label{eq:optical_depth}
    \tau_{\nu} =-\ln{\left(1-\frac{T_{\rm b}}{J_{\nu}(T_{\rm ex})-J_{\nu}(T_{\rm CMB})}\right)},
\end{equation}
with $J_{\nu}$ the radiation temperature. 

To compare our column densities derived from synthetic observations ($N_{\rm APEX}$ and $N_{\rm ALMA}$) to the model values ($N_{\rm Model}$), we integrated the o-H$_2$D$^+$ number density within the simulation box along the light-of-sight (LOS) and then derived $N_{\rm Model}$ by averaging the column densities over the area subtended by the beam. 
For the comparison of observed quantities with the model values, in the following sections we derive column densities over several angular scales, depending on telescope.
When computing column density ratios, the same angular scales were used for both the model and synthetic maps.


\begin{figure*}
    \centering
    \includegraphics[width=1.0\textwidth]{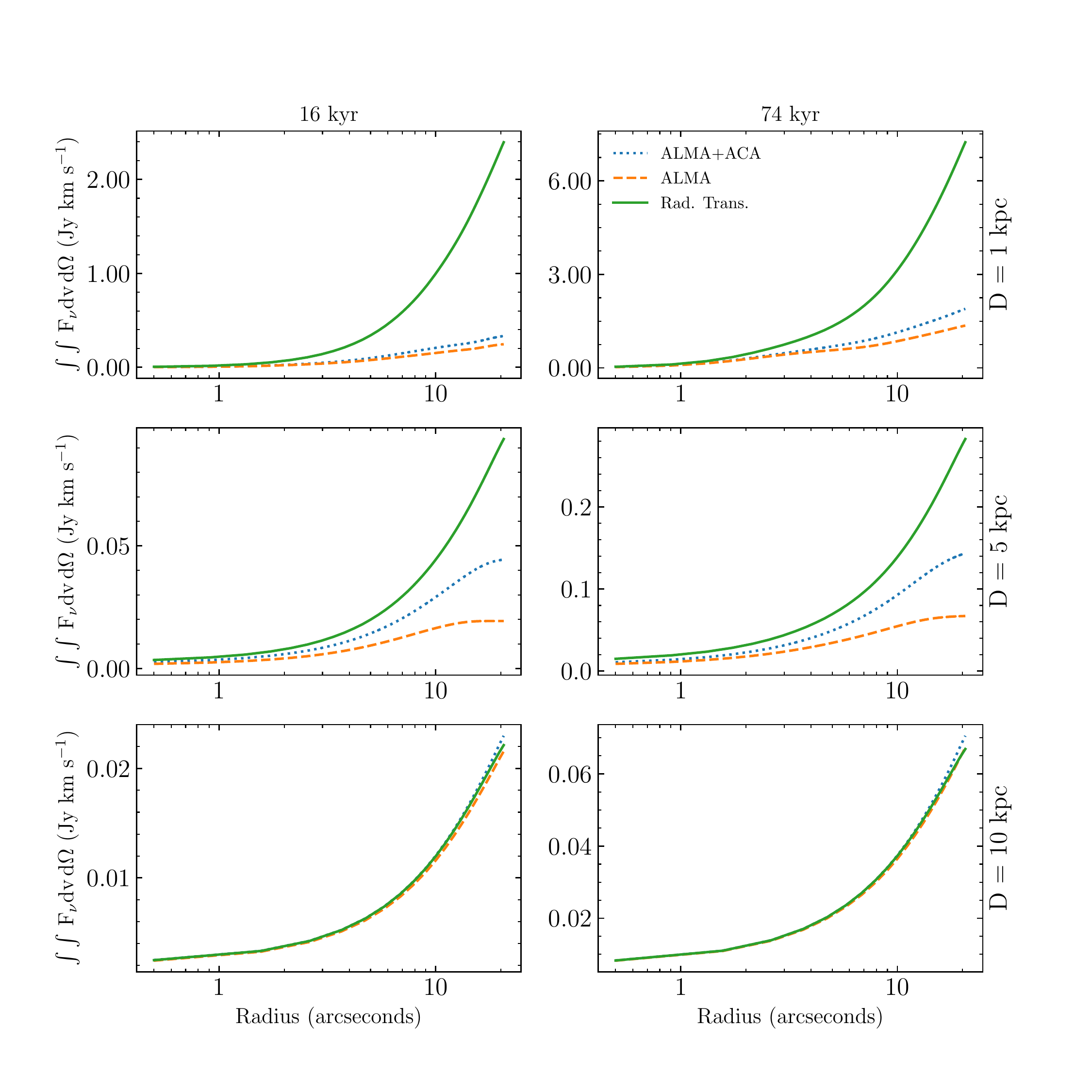}
    \caption{Cumulative distribution of the intensity at a distance of 1 kpc (top row), 5 kpc (middle row) and 10 kpc (bottom row), integrated within concentric circles of radii from 0.5 to 22 arcseconds, centered on the center of the map.  The distributions are shown for two evolutionary stages, at 16 kyr and 74 kyr, same as those depicted in Fig.~\ref{fig:mhd_snapshots} and~\ref{fig:radial_density_profiles}. The fluxes are shown for the output of the radiative transfer simulations (solid-green), for ALMA (dashed-orange) and for ALMA combined with ACA (dotted-blue). }
    \label{fig:radial_distributions_flux}
\end{figure*}

\begin{figure*}
    \centering
    \includegraphics[width=1.0\textwidth]{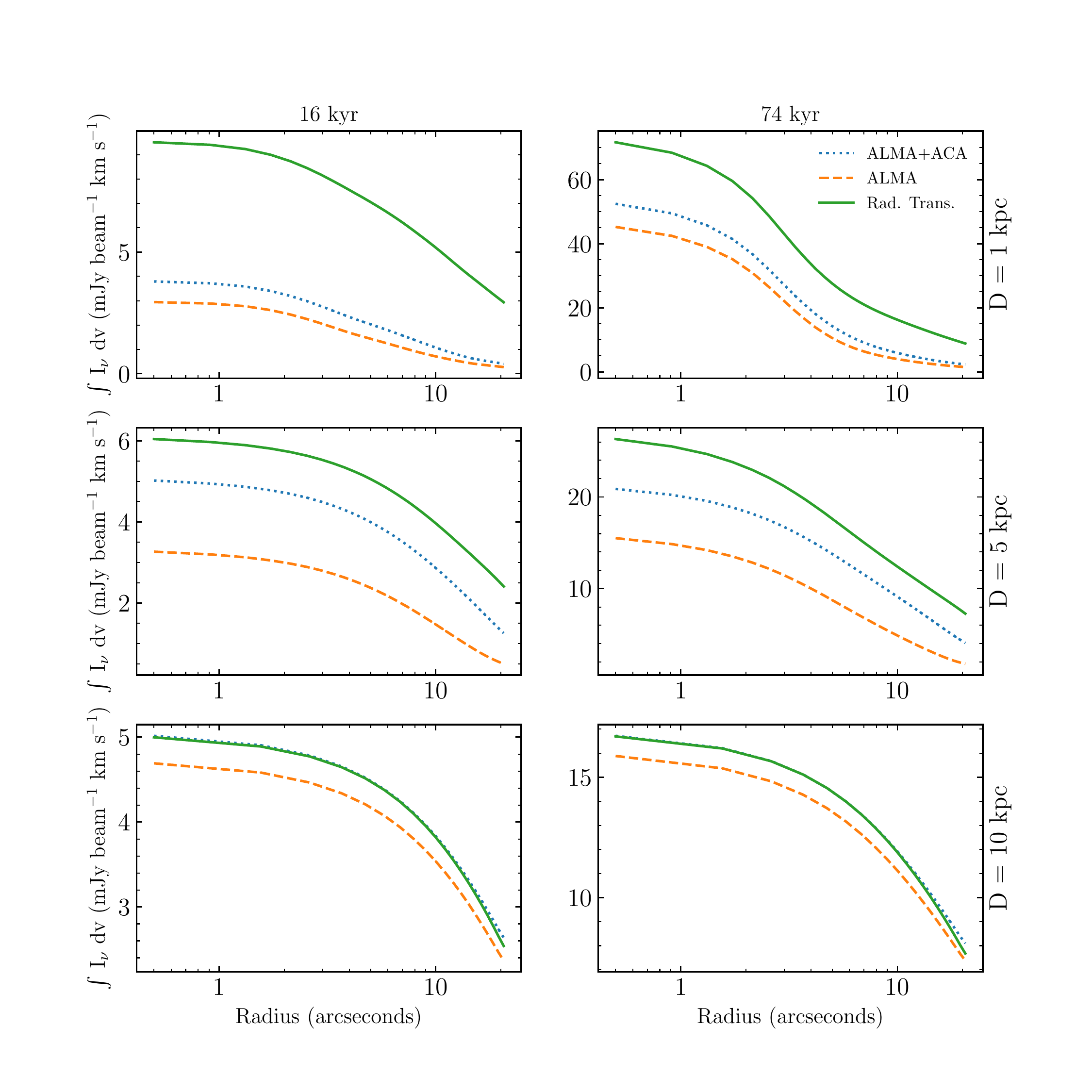}
    \caption{Radially averaged intensity distribution over the same scales presented in  figure~\ref{fig:radial_distributions_flux}.}
    \label{fig:radial_distributions_intensity}
\end{figure*}

\begin{figure*}
    \centering
    \includegraphics[width=1.0\textwidth]{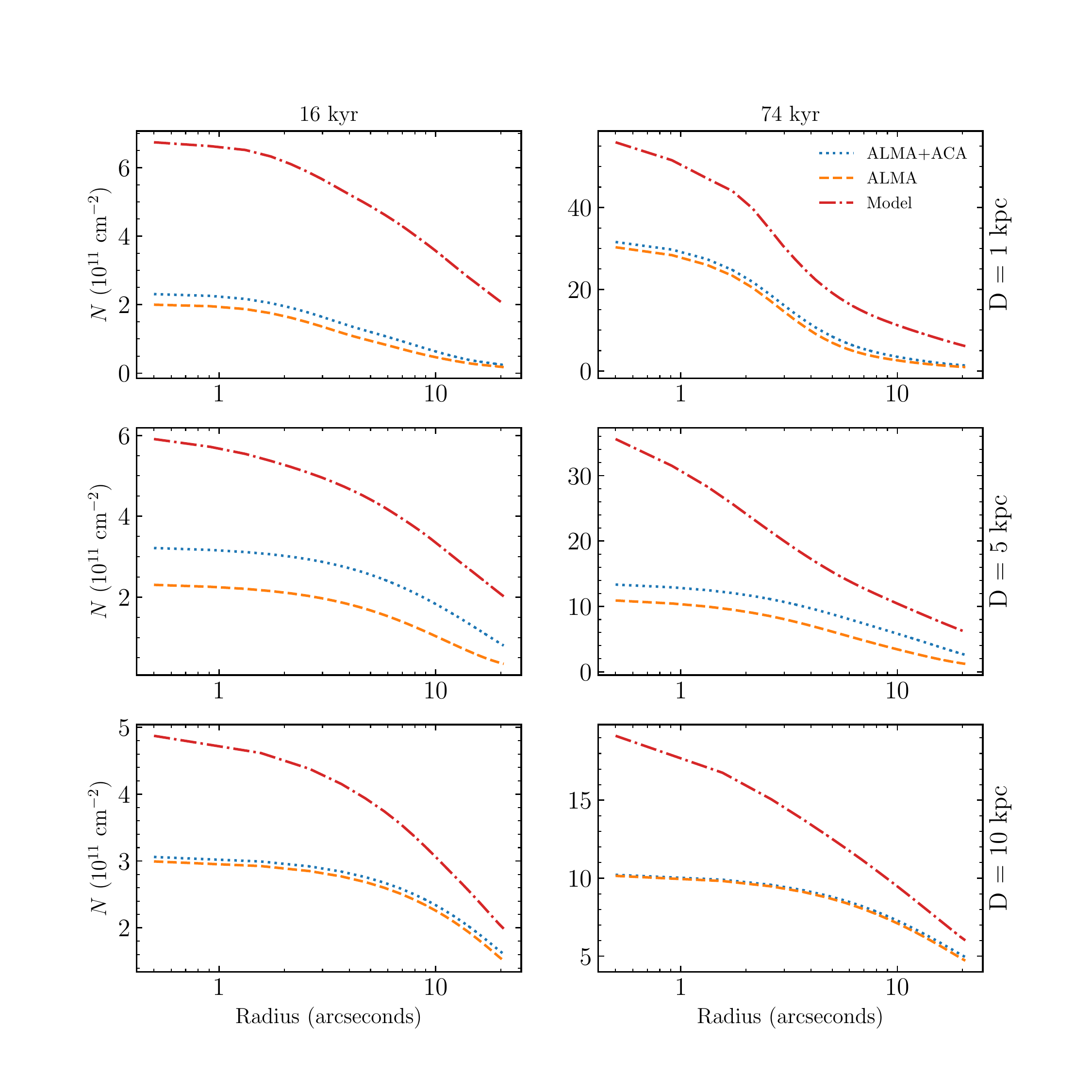}
    \caption{Radially averaged column density distributions over the same scales presented in figure~\ref{fig:radial_distributions_flux}. The model (semidashed-red) curve represent the physical column density integrated along the simulation box.}
    
    \label{fig:radial_distributions_N}
\end{figure*}

\section{Results}
\label{sec:Results}

\subsection{Radial distributions}
\label{subsec:Radial_distributions}
We first studied the fluxes, intensities and column densities as derived from the output of the RT simulation, then from the ALMA simulation and also ALMA combined with ACA.
The results presented in this section do not include thermal noise. 
This is to focus on the spatial and angular effects of the observations and to avoid any biased interpretation due to low signal-to-noise ratios (S/N).  
We performed observations using the two stages of the core shown in Fig.~\ref{fig:mhd_snapshots}, and we varied the source distance between 1, 5 and 10~kpc, to understand how the spatial filtering affects the retrieved column densities.
We show the resulting radially integrated flux distribution and radially averaged profiles of the intensity and column density, in Figs.~\ref{fig:radial_distributions_flux}, ~\ref{fig:radial_distributions_intensity} and \ref{fig:radial_distributions_N}, respectively. 
The two columns in each figure represent the stages at 16 and 74~kyr, from left to right. 
Each panel shows the results from the RT simulations in solid-green, from ALMA simulations in dashed-orange and from ALMA combined with ACA in dotted-blue. 
In Fig.~\ref{fig:radial_distributions_N}, we included the column densities obtained from the MHD simulation for reference, shown in semi-dashed-red.

The flux distribution in Fig.~\ref{fig:radial_distributions_flux} is an increasing function of radius because of the constantly increasing  area over which it is integrated, i.e., a cumulative function of intensity.
The increment is rather smooth for the early stage, at 16~kyr, since the gas surface density distribution is also a smooth function of radius, slightly evolved from the initial Bonnor-Ebert sphere. 
The interferometric observations show some relevant differences compared to the RT case, which are also higher at larger radii.
A possible explanation for this is that the largest angular scale of ALMA at the most compact configuration C43-1 (Band 7) is 8.25 arcseconds, while the size of the map is 41 arcseconds (at 1~kpc). 
Therefore, part of the extended emission is not detected by ALMA (main array only). 
This effect is slightly mitigated with the inclusion of ACA into the observing array (with a largest angular scale of 19.3 arcseconds), because it aids in covering the short-range baselines and allows to recover emission from larger angular scales, increasing the total fluxes and making them closer to those from the RT calculation. 
This effect is also appreciable when comparing the upper, middle and bottom panels, representing increasing distances of 1, 5 and 10\,kpc, respectively. 
The difference between fluxes from ALMA (both main array and combined) as compared to the RT case is largest at the closest distance and becomes almost negligible at 10\,kpc. 
The reason is the smaller angular scale that the source covers at larger distances, which allows all the extended emission to lie within the ALMA field of view. 
All these effects produced by the different spatial coverage are similarly reflected in the intensity distributions shown in Fig.~\ref{fig:radial_distributions_intensity}. 
The main difference in its radial profile as compared to that of the radial flux is that the intensity decreases for longer radii because it is computed as a beam averaged quantity.  
Since the large scale emission of o-H$_2$D$^+$ is very faint, the detectable emission is rather compact (as seen in Fig. \ref{fig:mhd_snapshots}), therefore averaging over larger beams certainly decreases the retrieved intensity.

The column densities were derived from the intensity using equation~(\ref{eq:column_density}) and therefore their radial distribution is similar to that of the intensity (see Fig.~\ref{fig:radial_distributions_N}). 
For the column density analysis, we also included the values derived from the MHD simulations, labeled $N_{\rm Model}$, which represent the true model values, in order to quantify the decrease in the measurement of column densities when derived from the ALMA simulations.  
The model values were obtained by integrating the o-H$_2$D$^+$ number density in the simulation box along the LOS and averaging it over concentric circles. 
The radii of such circles are the corresponding physical scales of apertures from 0.5 to $\sim$22 arcseconds, at distances of 1, 5 and 10~kpc.  
Our main purpose here is to check how well the column density can be recovered from the synthetic observation.
The differences between ALMA and the model are as high as in the case for the intensity and follow the same trend at both timesteps. 
The origin of these differences is the same as for the flux and intensity distributions. 

\subsection{Comparison of inferred o-H$_2$D$^+$ column densities from ALMA and APEX}
\label{subsec:Comparison_of_column_densities}
\begin{figure}
    \centering
    \includegraphics[width=1.0\columnwidth]{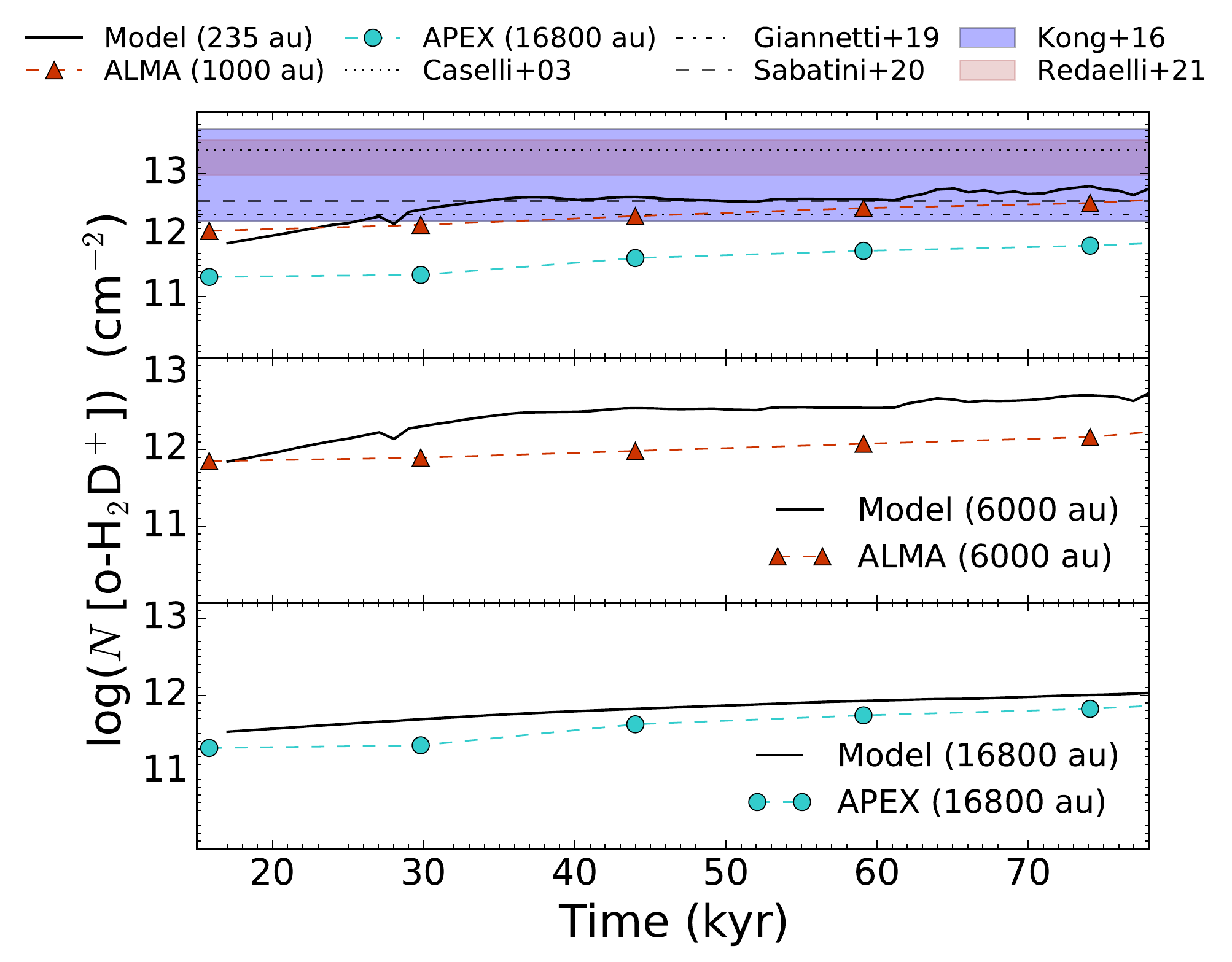}
    
    \caption[Model versus synthetic observations over collapsing time]{Model versus synthetic column densities as a function of the collapsing time, using a source distance of 1~kpc. The upper panel shows the column densities from the model, ALMA and APEX, within their maximum resolution element, e.g., 235~au, 1000~au (1") and 16800~au (16.8"), respectively. The middle panel shows the column densities from the model and ALMA derived both over the same extent of 6000~au (6"), which covers the densest and detectable parts of the core. The lower panel shows the model and APEX results, both averaged over the scale of covered by the APEX beam (16800~au).}
    \label{fig:column_densities_information_loss}
\end{figure}

We also compared our observed column densities to values reported in the literature, in order to asses the reliability of our model and framework for synthetic observations.  
Common values of the o-H$_2$D$^+$ column density toward low- and high-mass sources lie within 10$^{11}$ and 10$^{13}$~cm$^{-2}$ (see Table~\ref{tab:literature_values_of_column_density}). 
Most of these values come from single-dish observations, with the exception of the recent detections carried out with ALMA by \citet{Redaelli21}. 
Here we compare them to our results obtained from the single-dish and interferometric simulations, aiming to understand the key differences that arise when using any of the two approaches.
Our results are presented in Fig.~\ref{fig:column_densities_information_loss}.
We show the results of the model, for synthetic observations with ALMA and APEX, and for this analysis we present several timesteps over the core evolution, from 16\,kyr to 74\,kyr.
In the upper panel we also overplot the values from the literature, shown in Table~\ref{tab:literature_values_of_column_density}, for reference.
When comparing our results to real observations, we can see that, after $\sim$30\,kyr of evolution, the column densities derived from ALMA lie well within the range of values reported in the literature for single-dish observations \citep{Kong16,Giannetti19,Sabatini20}, but are lower than the results from \citet{Redaelli21} by a factor of 10.   
Similarly, the values derived for APEX are 6 times lower (on average) than the real APEX observations \citep{Giannetti19, Sabatini20}. 
The difference in the derived column densities is very likely related to the assumed value of the excitation temperature $T_{\rm ex}$, which we have fixed to 15\,K, under the assumption of full thermal coupling with the gas (i.e., LTE conditions). 
The survey of low-mass sources conducted by \citet{Caselli08} used a $T_{\rm ex}$ of $\sim$7\,K (average between all their starless cores), which was similarly used by \citet{Kong16} and slightly increased to $T_{\rm ex}=10$\,K by \citet{Redaelli21} when deriving column densities with ALMA. 
However, our ALMA synthetic column densities match better the results derived by \citet{Giannetti19} and \citet{Sabatini20}, who assumed  excitation temperatures in the range of $10 - 20$\,K for high-mass sources observed with APEX.   
To estimate the difference, we have additionally examined our column densities using lower excitation temperatures, and we find that a value of $T_{\rm ex}=7\,$K makes our ALMA values reproduce the observations from \citet{Redaelli21} and our APEX values reproduce those from \citet{Giannetti19} and \citet{Sabatini20}. 
As pointed out by \citet{Caselli08}, this comparison shows the effect that the selection of excitation temperatures has in the derivation of column densities. 
Within the overall uncertainties, we thus consider our theoretical expectations to be consistent with the observed fluxes. 

The upper panel of Fig. \ref{fig:column_densities_information_loss}, compares the column densities derived from the model, ALMA and APEX, over their respective highest resolution element, i.e., 235au, 1000\,au (1") and 16800\,au (16.8"), to see how relevant the limitation in spatial resolution is, and the retrieval of lower column densities due to the averaging over larger areas.  
As a proof of concept, the APEX values are lower than ALMA and the model because the emission of the core is very compact, and therefore the average intensity is lower for a larger beam. 
This means that for single-dishes and unresolved sources one must account for the beam filling factor of the source in order to avoid underestimations of the real column densities.

\begin{figure}
    \centering
    \includegraphics[width=1.0\columnwidth, trim=1.6cm 0.5cm 2.8cm 0, clip]{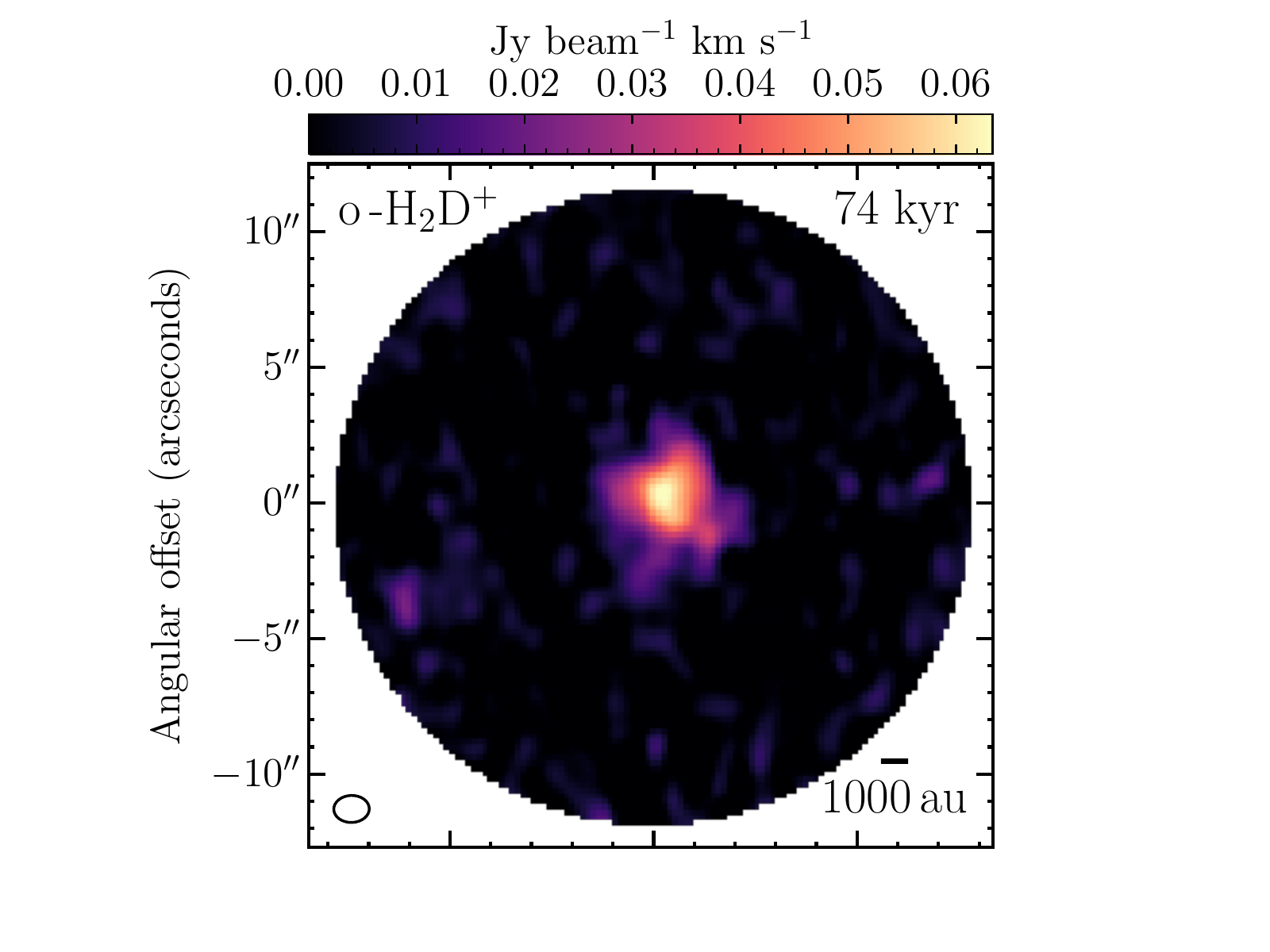}
    \caption{Integrated emission (moment 0) of o-H$_2$D$^+\,(1_{10}-1_{11})$ from an ALMA synthetic observation, produced by the core at the age of 74\,kyr (c.f., Fig. \ref{fig:mhd_snapshots}), at a distance of 1\,kpc and after observing for 10\,hrs.}
    \label{fig:mom0_oh2dp}
\end{figure}

In the second panel we show the model and ALMA values averaged over the same spatial extent, equal to 6000~au.  
This is about the largest angular scale obtained by ALMA in its most compact configuration and avoids the necessity to perform multiple pointings to cover the whole image. 
The scale is also sufficient to cover the densest regions that lead to detectable emission, as it can be seen in the synthetic ALMA map shown in Fig. \ref{fig:mom0_oh2dp}, for the core at the age of 74\,kyr after a 10\,hr observation.  
We compared over the same scale to focus on the effect of the sensitivity of the interferometer to the real physical values, without affecting the column densities by different spatial averages.
The third panel shows a similar comparison but for the case of the APEX resolution (16.8"), which covers 16800\,au.  
Based on the comparison in the middle and bottom panels, we see that when averaging over the same angular scale for both the model and APEX, the values retrieved by APEX are closer to the model than in the case for ALMA. 
This is due to the continuous spatial sampling that is achieved by single-dish telescopes, and points to the effect that the spatial filtering from interferometers has on decreasing the estimation of column densities. 

\begin{table}
    \caption[Column densities of o-H$_2$D$^+$ reported in the literature]{List of column densities of o-H$_2$D$^+$ (N[o-H$_2$D$^+$]) from cores and clumps reported in the literature and used in Fig.~\ref{fig:column_densities_information_loss}. }
    \label{tab:literature_values_of_column_density}
    \centering
    \setlength\tabcolsep{0.05pt}
    \begin{tabular}{lccl}
        \hline
        Telescope & Beam (") & $\log_{10}N$ (cm$^{-2}$) &   Source \\
        \hline
        CSO & 22    & 13.4 & Low-mass$^{a}$  \\
        CSO & 22    & 12.4 - 13.6 & Low-mass$^{b}$     \\
        JCMT & 15   & 12.3 - 12.7 & High-mass$^{c}$  \\
        JCMT & 15   & 12.2 - 13.7 & High-mass$^{d}$   \\
        APEX & 16.8 & 12.3 - 12.5 & High-mass$^{e}$ \\
        APEX & 16.8 & 12.1 - 13.0 & High-mass$^{f}$ \\
        ALMA & 1.00 & 13.0 - 13.5 & High-mass$^{g}$ \\
        \hline
        \multicolumn{2}{l}{$^{a}$\citet{Caselli03}} & \multicolumn{2}{l}{$^{b}$\citet{Caselli08}}  \\
        \multicolumn{2}{l}{$^{c}$\citet{Pillai12}} & \multicolumn{2}{l}{$^{d}$\citet{Kong16}}  \\ 
        \multicolumn{2}{l}{$^{e}$\citet{Giannetti19}} & \multicolumn{2}{l}{$^{f}$\citet{Sabatini20}}  \\
        \multicolumn{2}{l}{$^{g}$\citet{Redaelli21}} & & \\
        
    \end{tabular}
\end{table}

\begin{figure}
    \centering
    \includegraphics[trim=0.5cm 0.5cm 0.5cm 0.5cm, width=0.9\columnwidth]{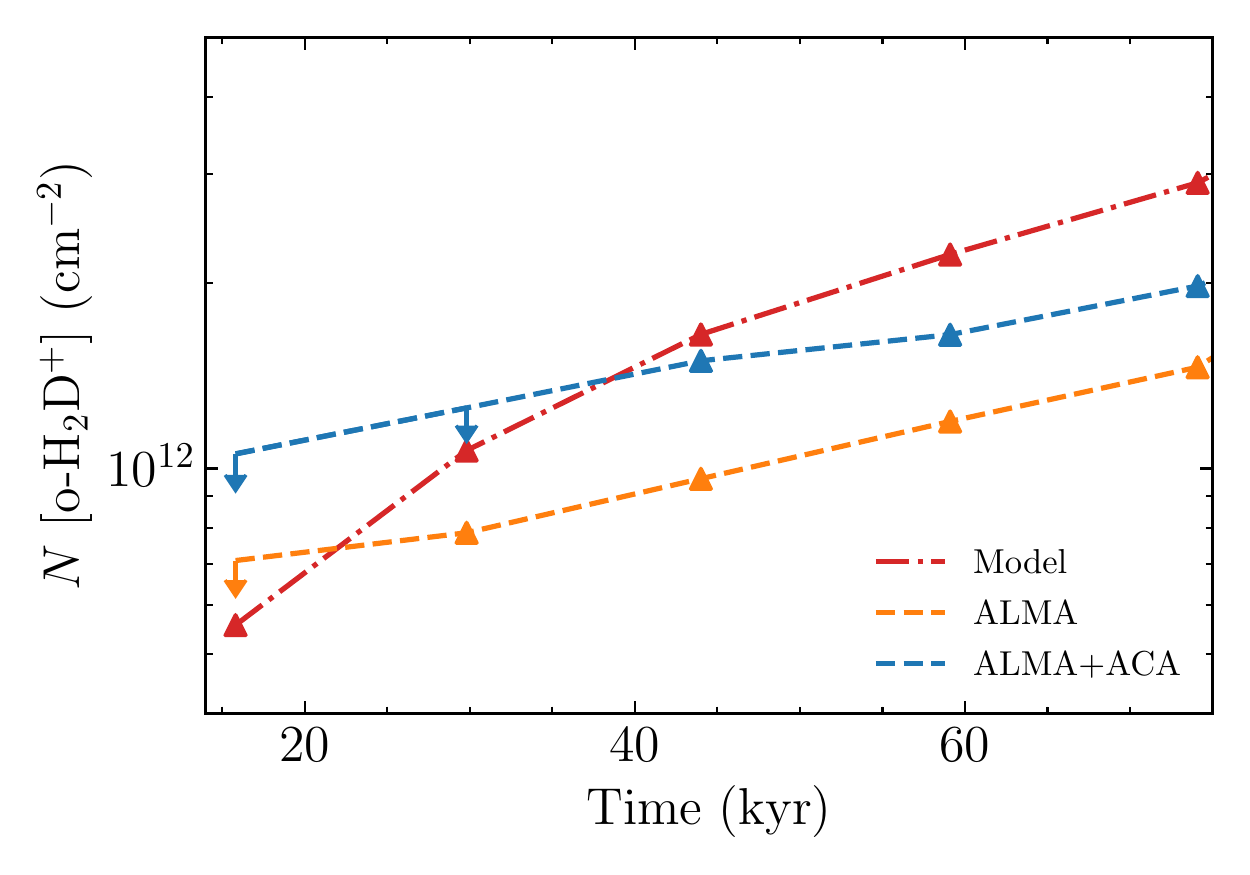}
    \caption{Column  density  evolution  over  time, analogous to Fig. \ref{fig:column_densities_information_loss}. All  points  are  averaged  within a beam of  6  arcseconds,  for  the  model  (semi-dashed red), for ALMA (orange) and for ALMA+ACA (blue). The downward arrows represent upper limits for the column density, since the detections at early timesteps is highly affected by the sensitivity of the observations.}
    \label{fig:column_densities_different_configs}
\end{figure}

We additionally reproduced the same analysis including the combined effect of ALMA and ACA, in order to see how significant the increase in the recovered flux is  when complementing the \textit{uv}-coverage with a sampling of the short-baselines. 
From the results shown in Fig.~\ref{fig:column_densities_different_configs}, we see that the inclusion of ACA increases the retrieved column densities at all stages of the collapse, similar to the effect seen in the radial profiles shown in Fig.~\ref{fig:radial_distributions_N}, but this time taking into consideration the sensitivity of the observations by the inclusion of thermal noise.   
For the first two stages of evolution, we present our results as upper limits on the column density because the low abundance of o-H$_2$D$^+$ at the early timesteps makes the emission to be highly affected by the sensitivity of the observations. 

\section{Discussion}\label{sec:Discussion}

\begin{figure}
    \centering
    \includegraphics[width=1.0\columnwidth]{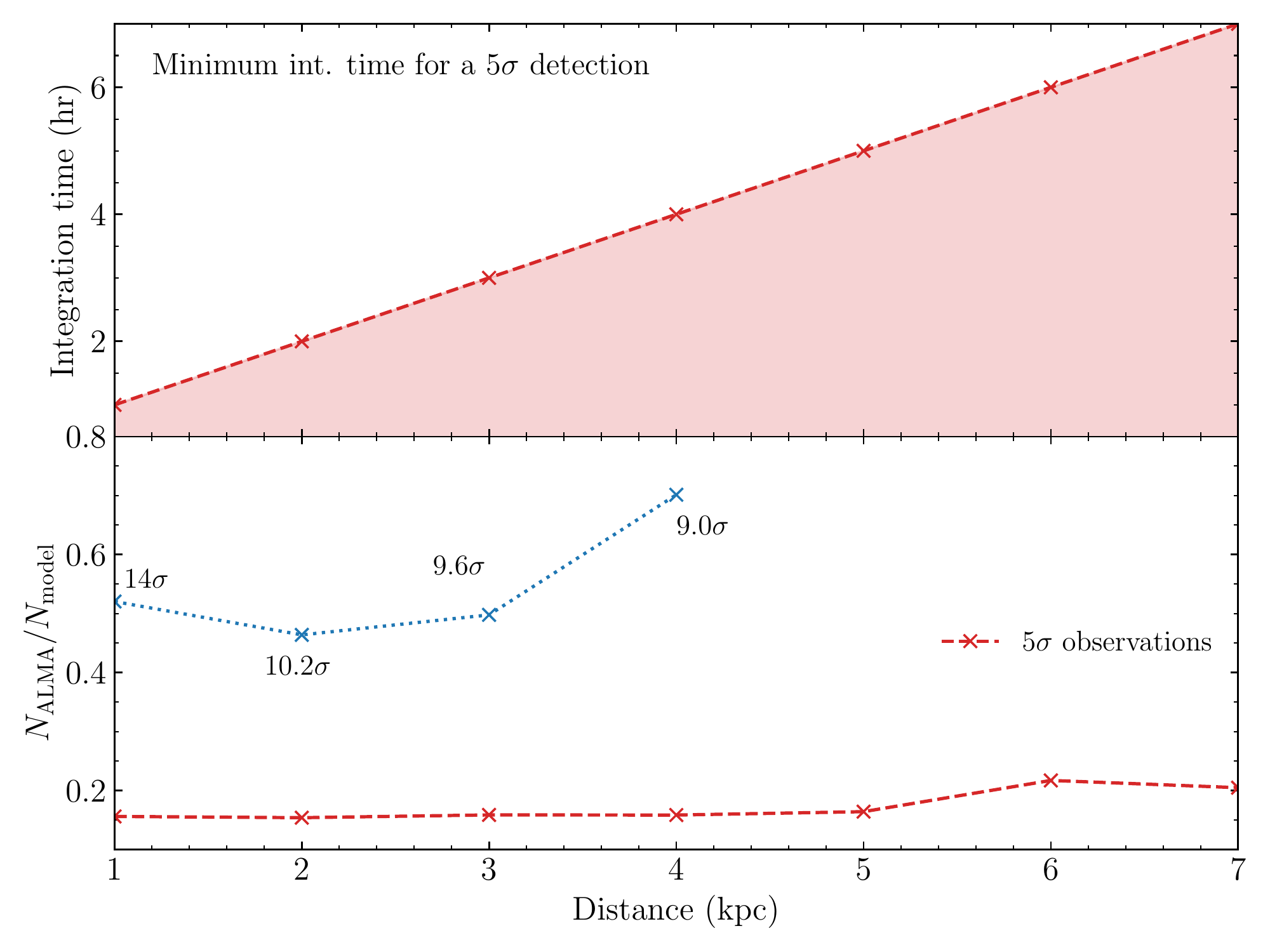}
    \caption{\textit{Top panel}: Minimum integration time required to have a 5$\sigma$ detection as a function of the distance to the source. The color-shaded region indicates observing times for which no detection was obtained. 
    \textit{Bottom panel}: Ratio of synthetic column densities as derived from ALMA over the values from the model, as a function of the source distance. The red line represent the ratios for the 5$\sigma$ observations and the blue line shows the results for high S/N observations with a 10\,hr integration time.}
    \label{fig:tint_vs_distance}
\end{figure}

\subsection{Minimum integration time for a $5\sigma$ detection}
\label{subsec:min_int_time}
We also studied the observability of the o-H$_2$D$^+$ $J_{\rm {K_a, K_c}} = 1_{10}$-$1_{11}$ line as a function of the distance to the source and the integration time required for interferometric observations only.
Here we are interested in the dependence of the minimum integration time required to have a 5$\sigma$ detection on the ratio of the column density measured over the model values when the source is placed farther away. 
For this analysis we performed ALMA (main-array only) simulations for the core at 74~kyr and at different distances, from 1 up to 7~kpc. 
The S/N value considered here was obtained by measuring the peak flux of the emission within the cube, in the regions and channels containing o-H$_2$D$^+$ emission (based on the ideal datacube from the radiative transfer) and dividing it by the cube rms.
The rms of each cube was obtained by deriving the mean rms over 4 rectangular regions far from the source, for a few channels far from the line emission and then averaging them all.
The results are shown in the upper panel of Fig.~\ref{fig:tint_vs_distance}. 
In order to keep the S/N constant at 5, we had to integrate longer for larger distances because of flux dependence on distance.  
At a distance of 1~kpc, a S/N of 5 can be reached after 1 hr of time on source.  
At a distance of 7~kpc, a S/N of 5 can be reached within at least 7 hrs of integration. 
All these simulations were binned along the spectral axis by a factor of 10, from $\Delta v=0.03$~km s$^{-1}$ to 0.3~km s$^{-1}$ to decrease the noise level, following equation (\ref{eq:T_rms_single-dish}).
 
Simulated observations lying on the red-shadowed region of Fig. \ref{fig:tint_vs_distance}, led to a S/N value lower than 5 and were classified as no detections.

\subsection{Quantifying the column densities information loss}
\label{subsec:Quantifying_the_underestimation_of_column_densities}

We are additionally interested in quantifying how much information is lost due to the post-processing presented in this paper. 
We estimate the amount of information loss by means of the ratio of column densities from the ALMA synthetic observations (with 1" resolution) and the column densities from the model (at intrinsic resolution), labeled $N_{\rm ALMA} / N_{\rm Model}$.  
We initially compute this ratio for the low S/N (5$\sigma$) observations presented in section \ref{subsec:min_int_time}. 
The results are shown in the lower panel of Fig.~\ref{fig:tint_vs_distance}, represented by the red data points.  
Our results indicate a constant underestimation of the column densities when the S/N is $\sim$5, happening for distances  up to 5\, kpc. 
This estimation seems to be slightly higher at 6 and 7\,kpc, likely produced by the decrease in the effect of spatial filtering, since all the emission of the source at 6 and 7\,kpc lies well within the ALMA beam.  
For these low-S/N observations, the model column densities are underestimated by around 85\% at 1~kpc and a bit more than 80\% at 7~kpc. 
These column density ratios are rather low, since the source emission that is detected above the noise limit is a tiny fraction of the actual emission. 

A natural question to ask is how much of the real column densities we lose if we aim for an actual high S/N detection. 
To understand this, we also performed ALMA observations at the same distances but increasing the integration time to 10~hrs each. 
The S/N of these observations were 14, 10.2, 9.6 and 9.0 for the 1, 2, 3 and 4~kpc distances, respectively, and calculated in the exactly the same way as in section \ref{subsec:min_int_time}.  
In Fig. \ref{fig:mom0_oh2dp} we present the resulting map of the integrated emission (moment 0) of o-H$_2$D$^+$, for the case of highest S/N observation, which is produced by the core at the age of 74\,kyr (from the density distribution shown in  Fig.\ref{fig:mhd_snapshots}), at a distance of 1\,kpc and observed for 10\,hrs. 
The results for the column density ratios at 1, 2, 3 and 4~kpc are shown by the blue line in the lower panel of Fig.~\ref{fig:tint_vs_distance}. 
For this part of the analysis, we only report the results up to 4~kpc because those at 5, 6 and 7~kpc did not achieve S/N much larger than 5, even when observed over several days or when binning down spectrally.
For the high-S/N observation at 1~kpc, the model column densities are underestimated by a factor of $\sim$0.46. 
For the 2~kpc case, the underestimation increases a bit to $\sim$0.59 and then decreases to $\sim$0.56 and $\sim$0.30 at 3 and 4~kpc.
The lack of a direct correlation between S/N and $N_{\rm ALMA} / N_{\rm Model}$ is due to the interplay between the spatial filtering and the low recovery of source emission at lower S/N.
In our results, the former is more relevant at 1 and 2~kpc while the latter is more relevant at 3 and 4~kpc.

\section{Summary \& Conclusions}
\label{sec:Summary_and_conclusion}
In this work we present a framework to produce synthetic observations using the radiative transfer code \textsc{Polaris} and the CASA observing tools.  
Our implementation focuses on the observability of the molecular line emission of ortho-H$_2$D$^+ J_{\rm {K_a, K_c}} = 1_{10}$-$1_{11}$ at 372.42~GHz generated from the simulation of an isolated, rather compact high-mass prestellar core. 
We studied the differences that may arise when deriving column densities from a physical model and when deriving them from the flux of a single-dish or interferometric observation.  
We present a proof of concept of our framework and show that it reproduces the observed fluxes. 
Our main conclusions are as follows: 
\begin{itemize}
    \item[-]
    Column density estimates directly depend on the size of the area over which they are averaged. 
    Then, when observing sources with compact emission, observations performed at lower angular resolutions will tend to have large losses due to beam dilution effects. 
    Similar results have been reported by \cite{Bovino19}.
    This is highly effective when comparing estimations between interferometric and single-dish observations.
    However, when comparing both approaches to the real values averaged over the respective spatial extent, single-dish telescope estimations are much closer to the model than from interferometers.
    This is the result of a lack of sensitivity to the more extended emission due to missing short baselines, which decreases the overall source sampling as compared to single-dish telescopes, where the sampling is more extended and continuous within the beam. 
    
    \item[-]
    The combined observations of ALMA and ACA improve the column density estimation with respect to the to the real values as compared to ALMA main array only, because it aids in reducing the effect of spatial filtering. 
    
    \item[-]
    The correlation between distance and the fraction of column density obtained by ALMA ($N_{\rm ALMA}/N_{\rm Model}$) is not linear.  
    Instead, it is determined by an interplay between the S/N and the spatial filtering. 
\end{itemize}

We emphasize that the best results to estimate the physical column densities from a source will be obtained through the combination of single-dish and interferometric observations. 
Observers commonly take this effect into account by including Total Power measurements in their setup. However, we have not included them in our analysis, to strictly compare single-dish versus interferometric observations.   
Our analysis confirms the effects that contribute to the information loss of the estimated column densities. 
These are, beam dilution in the case of single-dish telescopes and spatial filtering in the case of interferometers. 

\acknowledgements
JZ thanks for kind hospitality at Hamburg Observatory, the University of Kiel, and the Istituto di Radioastronomia (INAF-IRA) and Italian node of the ALMA Regional Centre (It-ARC) of Bologna.
This research made use of NASAs Astrophysics Data System Bibliographic Services (ADS), of Astropy, a community-developed core Python package for Astronomy (\citealt{astropy2013,astropy2018}; see also http://www.astropy.org), Matplotlib~\citep{Hunter07} and APLPy, an open-source plotting package for Python \citep{aplpy2012,aplpy2019}.


\begin{fundinginformation}
JZ and DRGS thank for funding via Fondecyt regular 1161247 and 1201280. 
SB and DRGS gratefully acknowledges support by Conicyt Programa de Astronomia Fondo Quimal 2017 QUIMAL170001 and by the ANID BASAL project FB210003.
SB thanks for funding through Fondecyt Iniciacion 11170268. 
SB and JZ also thanks for funding through the DFG priority program “The Physics of the Interstellar Medium” (project BO 4113/1-2).
The simulations were performed with resources provided by the \emph{Kultrun Astronomy Hybrid Cluster} at the Department of Astronomy, Universidad de Concepci\'on. 
\end{fundinginformation}

\begin{dataavailability}
The datasets generated during and/or analysed during the current study are available from the corresponding author on reasonable request.
\end{dataavailability}

\begin{codeavailability}
The code used in this paper is available at \url{https://github.com/jzamponi/synthetic_module}.
\end{codeavailability}

\begin{authorcontribution}
All authors contributed to the study conception and design. Material preparation, data collection and analysis were performed by all authors. The first draft of the manuscript was written by Joaquin Zamponi and all authors commented on previous versions of the manuscript. All authors read and approved the final manuscript.
\end{authorcontribution}

\begin{ethics}
\begin{conflict}
None.
\end{conflict}

\smallskip
\noindent\textbf{Informed consent} None.

\end{ethics}


%
\bibliographystyle{spr-mp-nameyear-cnd}  
\bibliography{references}                

\end{document}